\documentclass[journal,onecolumn,12pt]{IEEEtran}
\usepackage{comment}

\usepackage{mathptmx}
\usepackage{geometry}

\usepackage{ifpdf}

\usepackage{algorithmicx}
\usepackage{times}
\usepackage{fancyhdr,graphicx,amsmath,amssymb}
\usepackage[ruled,vlined]{algorithm2e}
\usepackage{cite}
\include{pythonlisting}
\makeatletter
\def\BState{\State\hskip-\ALG@thistlm}
\makeatother

\usepackage{cite}
\usepackage{amsmath}
\usepackage{amssymb}
\usepackage{flexisym}
\usepackage{array}
\newcolumntype{C}[1]{>{\centering\let\newline\\\arraybackslash\hspace{0pt}}m{#1}}

\usepackage{color}
\definecolor{dblue}{rgb}{0,0,0.8}

\usepackage{xcolor}

\usepackage{multirow}

\usepackage{subfig}
\usepackage{graphicx}
\usepackage{fixltx2e}

\usepackage{stfloats}
\usepackage{subfig}

\usepackage[utf8]{inputenc}

\usepackage{amssymb}

\usepackage{nomencl}
\makenomenclature
\usepackage{etoolbox}
\usepackage{hyperref}

\usepackage{url}

\usepackage{mhchem}

\usepackage{nomencl}
\makenomenclature

\usepackage[justification=centering]{caption}

\usepackage{nomencl}
\makenomenclature

\usepackage{hyperref}

\begin{document}
%
\title{Hydrogen and Battery Storage Technologies for Low Cost Energy Decarbonization in Distribution Networks}%
%
%

\author{\normalsize{Hamed~Haggi, Paul Brooker,
        Wei~Sun, and James M. Fenton}
\thanks{This work is supported in part by U.S. Department of Energy's award under grant DE-EE0008851. H. Haggi and W. Sun are with the Department of Electrical and Computer Engineering, University of Central Florida, Orlando, FL 32816 USA (e-mail: hamed.haggi@knights.ucf.edu, sun@ucf.edu). J. M. Fenton is with University of Central Florida's FSEC Research Energy Center, Cocoa, FL 32922 USA (jfenton@fsec.ucf.edu) and P. Brooker is with Orlando Utilities Commission, Orlando, FL 32839 USA (PBrooker@ouc.com).}}

\maketitle

\begin{abstract}
 Deep energy decarbonization cannot be achieved without high penetration of renewables. At higher renewable energy penetrations, the variability and intermittent nature of solar photovoltaic (PV) electricity can cause ramping issues with existing fossil fuel generation, requiring longer term energy storage to increase the reliability of grid operation. A proton exchange membrane electrolyzer can produce $H_2$ and serves as a utility controllable load. The produced $H_2$ can then be stored and converted back into electricity, or mixed with natural gas, or used as transportation fuel, or chemical feedstock. This paper considers the perspective of the distribution system operator that operates the distributed energy resources on a standard IEEE 33-node distribution network considering the technical and physical constraints with the goal of minimizing total investment and operation cost. Different case studies, at very high PV penetrations are considered to show the challenges and path to net-zero emission energy production using $H_2$ energy.  Sensitivity of utility PV costs and electrolyzer capital costs on producing $H_2$ at \$1/kg are presented showing that the distribution network could produce 100\% renewable electricity and $H_2$ could be produced at a cost of \$1/kg by 2050 with conservative cost estimates and by 2030 with accelerated cost declines.
\end{abstract}

\begin{IEEEkeywords}
 Deep Energy Decarbonization, Hydrogen Production Cost, Hydrogen System, Li-ion Battery, Vanadium Redox Flow Battery.  
\end{IEEEkeywords}

%
\IEEEpeerreviewmaketitle

\section{Introduction}

Population growth and transportation electrification will significantly increase electricity demand, and a continued reliance on fossil fuels for supplying this demand will contribute to the global warming crisis. According to the U.S. Energy Information Administration, electricity production accounts for 33\% of annual carbon dioxide ($CO_2$) emissions of which 65\% of this amount is produced by large-scale coal and natural gas power plants \cite{USEIA}. As a consequence, the global temperature is estimated to rise more than $3\, ^\circ$C on average by the year 2050. To that end, the U.S. government recently set a nationwide Net-Zero Emission goal by the year 2050 in which the power sector, especially power utilities and generation companies must supply their load with green energy by 2035, and the transportation sector must be electrified completely by 2050 \cite{IEA}. Additionally, the advancements in distributed energy resource (DERs) technologies, such as solar, wind, storage, hydrogen ($H_2$), energy storage systems, and demand response programs, will lower the capital costs of these technologies by 60\%-70\% by 2050 according to the National Renewable Energy Lab’s (NREL) Advanced Technology Baseline (ATB), further accelerating the growth in DERs. \cite{Vimmerstedt2018ATB}. Therefore, utilities hope to deploy more DERs to maximize green energy production while minimizing the total investment and operation cost compared to today's power systems. Large deployment of DERs, can control the world's temperature rise to less than $2\, ^\circ$C.\par

In recent years, $H_2$ used as an energy carrier or reducing agent, has great potential for decarbonization of both electricity production and manufacturing. $H_2$ can be produced by electrolyzers in a water electrolysis process. There are various types of electrolyzers for water electrolysis such as alkaline, proton exchange membrane (PEM), and solid oxide electrolysis, in which the electricity coming from renewable energy or grid is consumed by electrolyzers to decouple water  into $H_2$ and oxygen ($O_2$). $H_2$ can be stored in the form of cryogenic-liquid or pressurized gas and it can be used for 1) power generation with stationary/mobile fuel cell (FC) units \cite{haggi2022proactiveee}; 2) transportation fuel for light- and heavy-duty FC vehicles; 3) fuel for residential and commercial buildings (space and water heating); 4) chemical feedstocks i.e. ammonia production \cite{palys2020using}, etc., 5) utility control of the electrolyzer load and fuel cell output to allow solar photovoltaic (PV) smoothing, load shifting, voltage, and frequency regulations, etc.; and 6) mixed with natural gas to produce electricity through gas turbines \cite{kovavc2021hydrogen}\cite{roadmap}.\par

In addition to the completed and ongoing $H_2$ scale projects which are comprehensively reviewed in \cite{egeland2021hydrogen}, there are a large number of research efforts focusing on different aspects of $H_2$ system optimization such as modelling, chemistry and material research. This paper mainly focuses on the long-term operation scheduling of these systems. The impacts of power to gas technologies for decarbonization concerns were investigated in \cite{li2021modelling} by considering a coordinated planning between power and transportation systems. A modelling and operation strategy for hybrid systems including battery and $H_2$ energy storage integrated with wind farms was investigated in \cite{wen2020research}. Centralized scheduling of distributed $H_2$ stations under dynamic $H_2$ pricing was proposed in \cite{el2018hydrogen} considering the capacity-based demand response (DR). A supervisory-based framework for operation scheduling of $H_2$ refueling stations is proposed in \cite{khani2019supervisory}, in which the central entity (operator of hydrogen refueling stations) uses the total reserved capacity to assist in grid operation. In \cite{farag2020optimal}, optimal operation management for centralized and distributed electrolysis-based $H_2$ generation and storage systems was developed to maximize the net revenue of private investors. In \cite{xiao2010hydrogen}, an operating cost minimization model was proposed for scheduling $H_2$ production to meet the $H_2$ demand of fuel cell electric vehicles. An economic feasibility study of selling $H_2$ energy in the $H_2$ market was investigated in \cite{taljan2008feasibility} with the $H_2$ storage integrated with the power grid. The techno-economic analysis and feasibility study of wind power plants including $H_2$ storage were investigated in \cite{eypasch2017model}. Another design framework was proposed in \cite{blazquez2019techno} considering the best design for a $H_2$ fueling station, in terms of the number of tank banks, size, and cost. Moreover, optimal scheduling of $H_2$ storage to minimize cost and emission was studied in \cite{seyyedeh2019optimal} considering DR and renewable energies. An optimal scheduling of an energy hub using $H_2$ storage considering thermal demand response is proposed in \cite{alizad2020optimal}. A decentralized local energy market for electricity and $H_2$ trading was proposed in \cite{xiao2017local} through a multiplayer game-based market clearing algorithm with the privacy preservation of market participants. Stochastic scheduling of $H_2$ storage with wind and DR was investigated in \cite{mirzaei2019stochastic, yu2019risk}. A risk based analysis for onsite PV and $H_2$ system operation in active distribution networks using demand response was investigated in \cite{Haggi2021risk}. The authors of \cite{wu2020cooperative}proposed a distributed cooperative framework for wind and $H_2$ refueling stations' operation using Nash bargaining theory, where the uncertainties of wind power and electricity price were considered.\par

\begin{figure}
\centering
\includegraphics[width=4.5in]{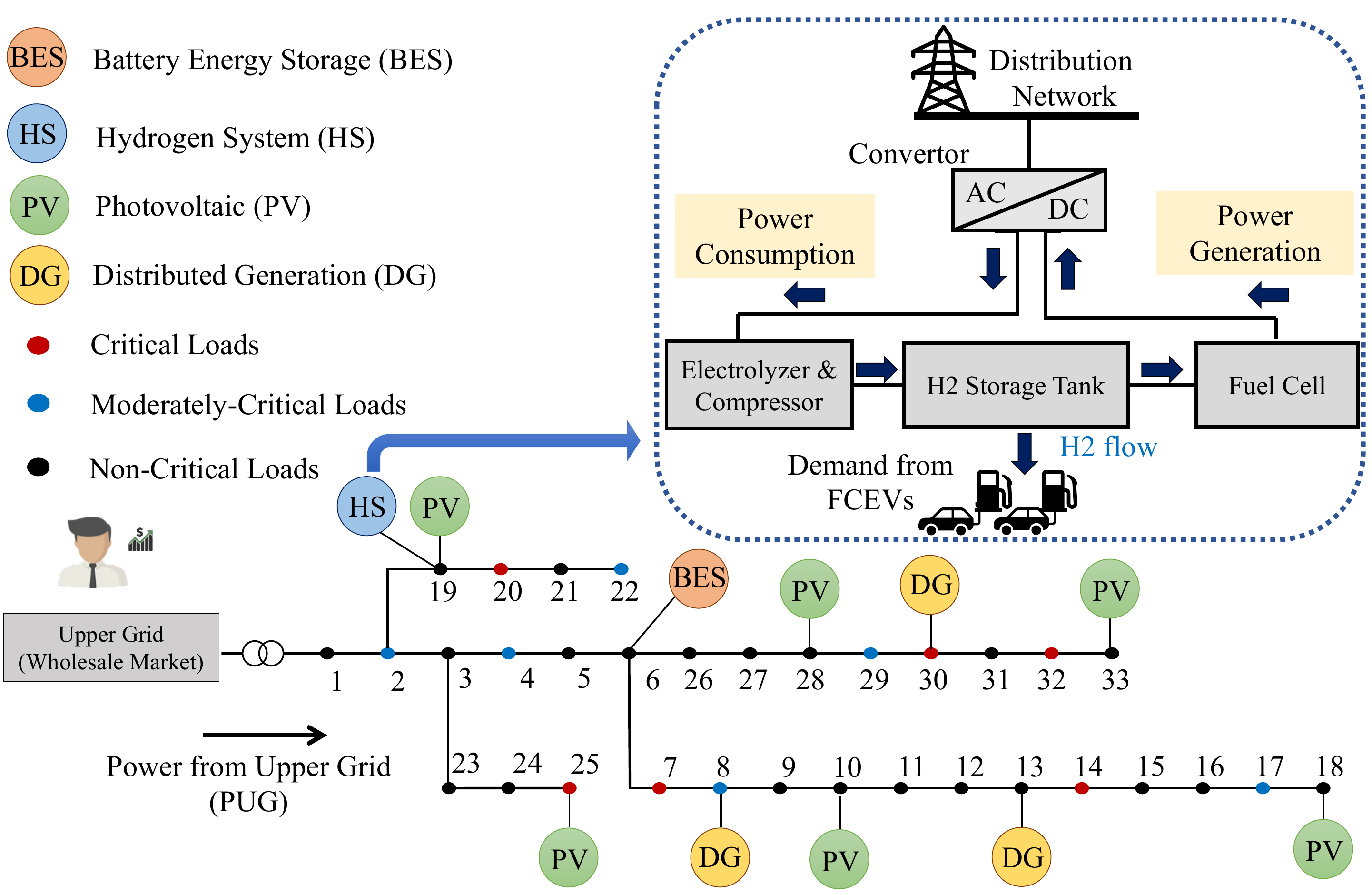}
	\caption{33-node active distribution network considering DERs.}
	\label{system}
\end{figure}

In this paper, we consider the perspective of the distribution system operator (DSO) that manages the natural gas power plants (distributed generators, DGs), the high penetration of solar PV energy, battery storage and $H_2$ production and consumption by stationary FCs, to reach the goal of net-zero emission energy production. To have realistic analysis of the vertically integrated distribution network considering the power flow and voltage challenges, a standard 33-node distribution network \cite{baran1989network} based on Fig. \ref{system} was studied. This network includes utility-operated natural gas power plants (combined cycle units and combustion turbine units), PV units, battery energy storage (Li-ion batteries, 10-hour duration Vanadium Redox flow batteries, etc.), and $H_2$ system (including electrolyzers, compressors, storage tanks, and stationary PEM FC units). Different types of voltage-dependent loads are considered such as critical, moderately-critical and non-critical loads to resemble load types like hospitals, offices, grocery stores, etc. The goal of normal operation from DSO's perspective is to operate these assets to minimize the total operational and investment costs and maximize the green energy production for the power sector. More details regarding the modeling can be found in \cite{haggi2021proactive}. \par

\section{Proposed Case Studies}
A path to Net-Zero Emissions is described using the following case studies (all of which use 2050 equipment and levelized cost of electricity) which show the importance of deploying DERs in the distribution networks. For this paper no electric vehicles nor fuel cell vehicles are included in the following case studies.
\begin{itemize}
    \item Case 1: Optimal system operation based on natural gas power plants (defined as DG).
    \item Case 2: Optimal system operation based on DGs and PVs.
    \item Case 3: Optimal system operation based on DGs, PVs, and 4-hour duration Li-ion battery.
    \item Case 4: Optimal system operation based on DGs, PVs, and 10-hour duration redox flow battery.
    \item Case 5a: Optimal system operation based on DGs, PVs, and long-term $H_2$ storage.
        \item Case 5b: Optimal system operation based on DGs, PVs, and long-term $H_2$ storage with updated LCOEs of DGs.
    \item Case 6a: Optimal system operation based on DGs, PVs, 4-hour duration Li-ion battery and long-term $H_2$ storage.
        \item Case 6b: Optimal system operation based on DGs, PVs, 4-hour duration Li-ion battery and long-term $H_2$ storage while penalizing curtailed PV energy. 
    \item Case 7a: Optimal system operation based on DGs, PVs, 10-hour duration redox flow battery and long-term $H_2$ storage with 10\% initial and final level of $H_2$ in the tank. 
        \item Case 7b: Optimal system operation based on DGs, PVs, 10-hour duration redox flow battery and long-term $H_2$ storage with 50\% initial and final level of $H_2$ in the tank. 
    
\end{itemize}

More information such as green energy production based on total electricity generation, comparison of DGs' capacity factor, the electricity cost delivered to customers (known as distributional locational marginal price (DLMP)), total PV capacity and the amount of curtailed PV energy will be presented for all the above mentioned case studies. Additionally, sensitivity analysis results will be presented for $H_2$ production cost including the electrolysis and storage cost with various capital expenditures (CAPEX) of electrolyzer and PV. The LCOE and CAPEX cost related parameters of each energy asset are presented in Table \ref{appendix}. Moreover, voltage values for each node during the two week period optimization horizon will be shown for Case 7a. For a comparison, Case 2 voltage values are shown to highlight the role of DERs in voltage regulation.

\section{Simulation Results and Analysis}
The IEEE 33-node distribution network \cite{baran1989network} shown in Fig.\ref{system} is considered for operation scheduling of assets such as DGs, PVs, $H_2$ system, and battery energy storage technologies such as Li-ion and vanadium redox flow. The $H_2$ system includes electrolyzers, compressors, storage tanks, stationary PEM FC units, and an inverter. The system hosts three DG units (located at nodes 8, 13, and 30 \cite{haggi2022proactiveee}), six PV units (located at nodes 10, 18, 19, 25, 28, 33) based on their operational advantages such as dispatched PV energy and voltage regulation constraints, one $H_2$ system at node 19 and one battery storage at node 6. The optimal location and sizes of assets that were added to the system such as electrolyzer, storage tank capacity, FC, Li-ion and redox flow batteries were determined based on the capital costs, lifetime, interest rate, operational advantages such as voltage improvements to minimize the total system cost (including both operation and capital costs). \par

In this paper, the DG located at node 8 is a natural gas combined cycle unit operating with the capacity of 0.8 MW and operational cost of \$36/MWh. DGs located at nodes 13 and 30 are natural gas combustion turbine units operating with the capacities of 2.4 MW and 1 MW, respectively \cite{haggi2022proactiveee}. The operational cost of these DGs are \$95/MWh, and \$98/MWh, respectively. It should be noted that the operational costs are chosen based on the NREL ATB file \cite{Vimmerstedt2018ATB} for the year 2050 assuming a capacity factor of 88\%, 12\%, and 12\% for DG8, DG13, and DG30 respectively. More information regarding the power plant parameter identification can be found in \cite{khazeiynasab2021power} \cite{khazeiynasab2021generator}. In addition to the cost related parameters, ramp rate limits as well as fixed no-load costs of DGs were considered. This information is available in our previous work \cite{Haggi2021risk}. In this paper, the goal is to minimize the total operation and investment cost considering the physical network and operational constraints of DGs, PVs, batteries, and $H_2$ storage. 

\begin{figure}[]
\centering
\footnotesize

	\includegraphics[width=4.5in]{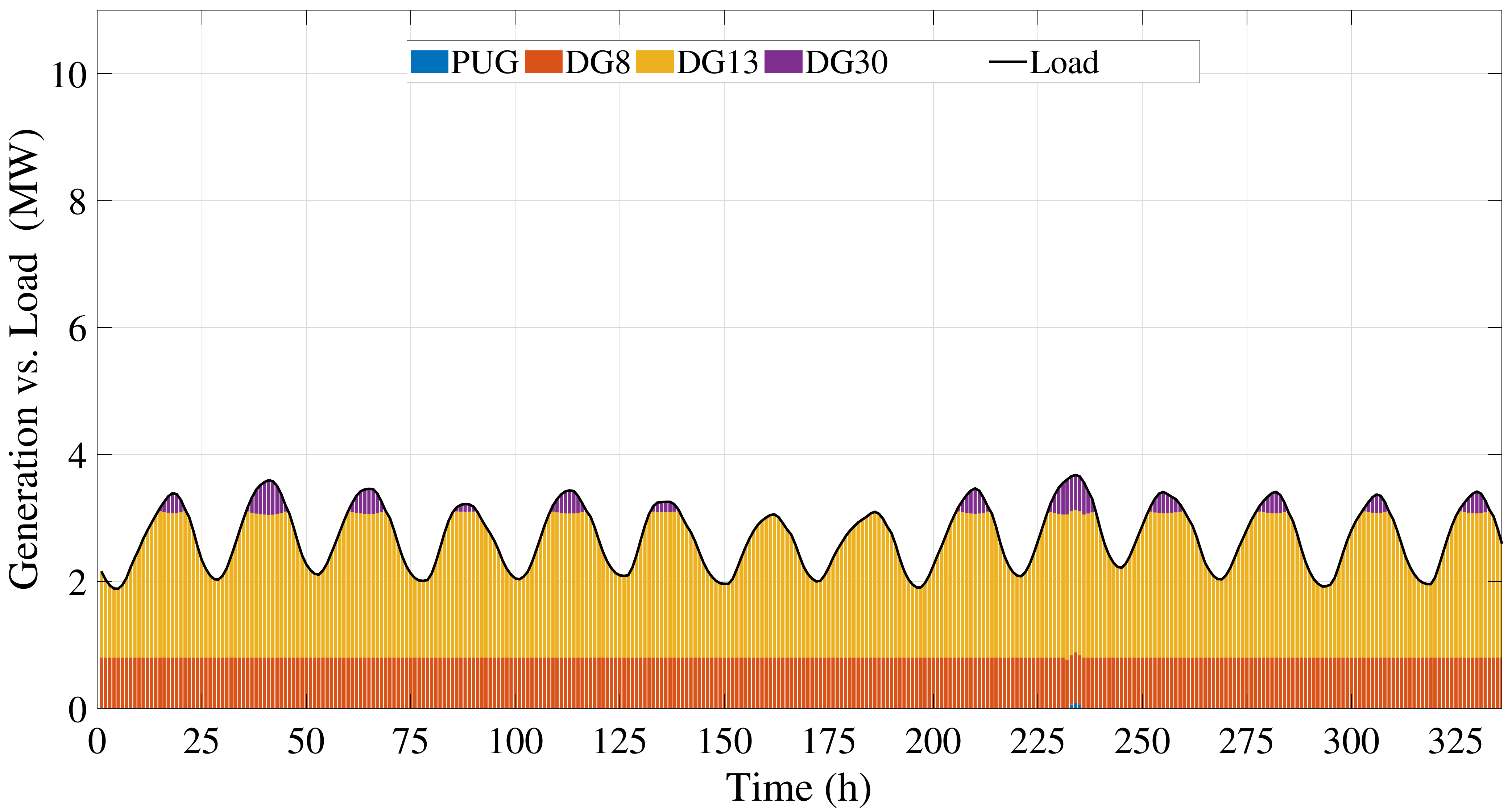}
	\caption{Optimal system operation based on DGs (Case 1).}
	\label{Case1}
\end{figure}

\subsubsection{Optimal system operation based on DGs (Case 1)} 
Fig.\ref{Case1} shows the results for when the DSO supplies the electricity demand of customers (known as grid load and shown by the black line) which is 925 MWh for the two week period only which is provided by the three DG units and by purchasing power from the wholesale market, shown as PUG in Fig.\ref{system}, (cost of \$100/MWh). In vertically integrated distribution test systems the DSO prefers to operate their own assets in the distribution network and not to purchase power from the upper grid (wholesale market). Fig. \ref{Case1} shows the simulation results for Case 1 in which three DGs supply the electricity demand of the customers except for the time period of 234 to 236 hours in which the DSO must purchase some power from the upper grid (PUG), shown with blue columns at the base of the figure, to handle the physical power flow and keep the voltages within the reliable operation range. Since the overall goal is to minimize the cost, DG8 (shown with orange columns) is operated first since the operational cost is lower than other two DGs. More information can be found in \cite{Haggi2021risk} and \cite{haggi2021proactive}.

\subsubsection{Optimal system operation based on DGs and PVs (Case 2)} 
\begin{figure}
\centering
\footnotesize
	\includegraphics[width=4.5in]{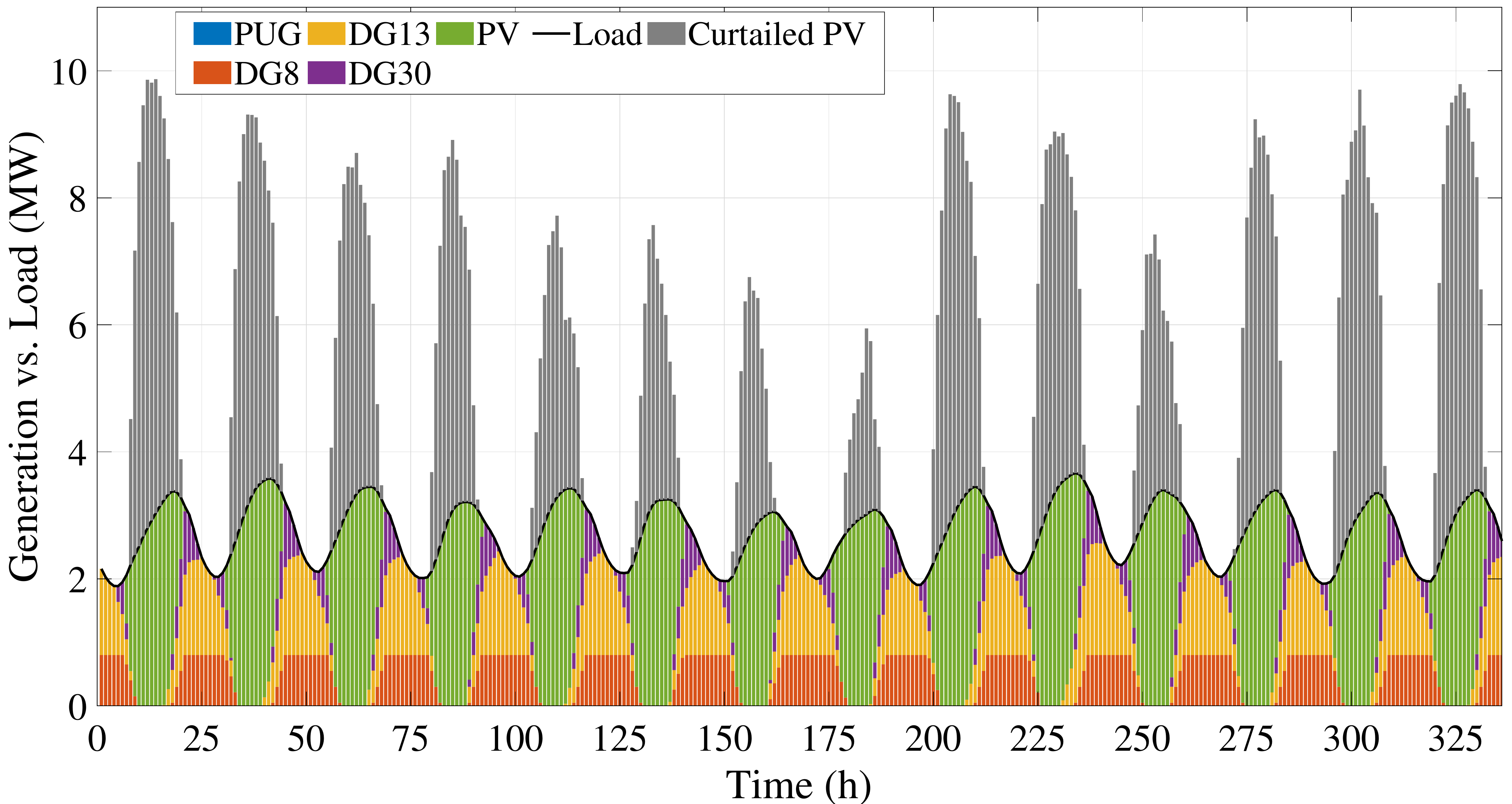}
	\caption{Optimal system operation based on DGs and PVs (Case 2)}
	\label{Case2}
\end{figure}

Compared to the previous case, six PV units (with levelized cost of energy of \$12/MWh with an assumed capacity factor of 33\%) are added to the system at the locations shown in Fig.\ref{system}. Fig. \ref{Case2} shows the generation vs. load curve for high PV penetration levels (120\% of total grid load which equals to 1110 MWh for the two weeks period). The PV units supply the grid load when solar energy is available. However, inclusion of PV energy causes an increase in DG30 (shown as purple columns) operation due to the ramp rate constraints of the DGs. The surplus PV energy that exceeds the load is curtailed which is shown by the grey columns.

\subsubsection{Optimal system operation based on DGs, PVs, and 4-hour duration Li-ion battery (Case 3)} 
\begin{figure}
\centering
\footnotesize

	\includegraphics[width=4.5in]{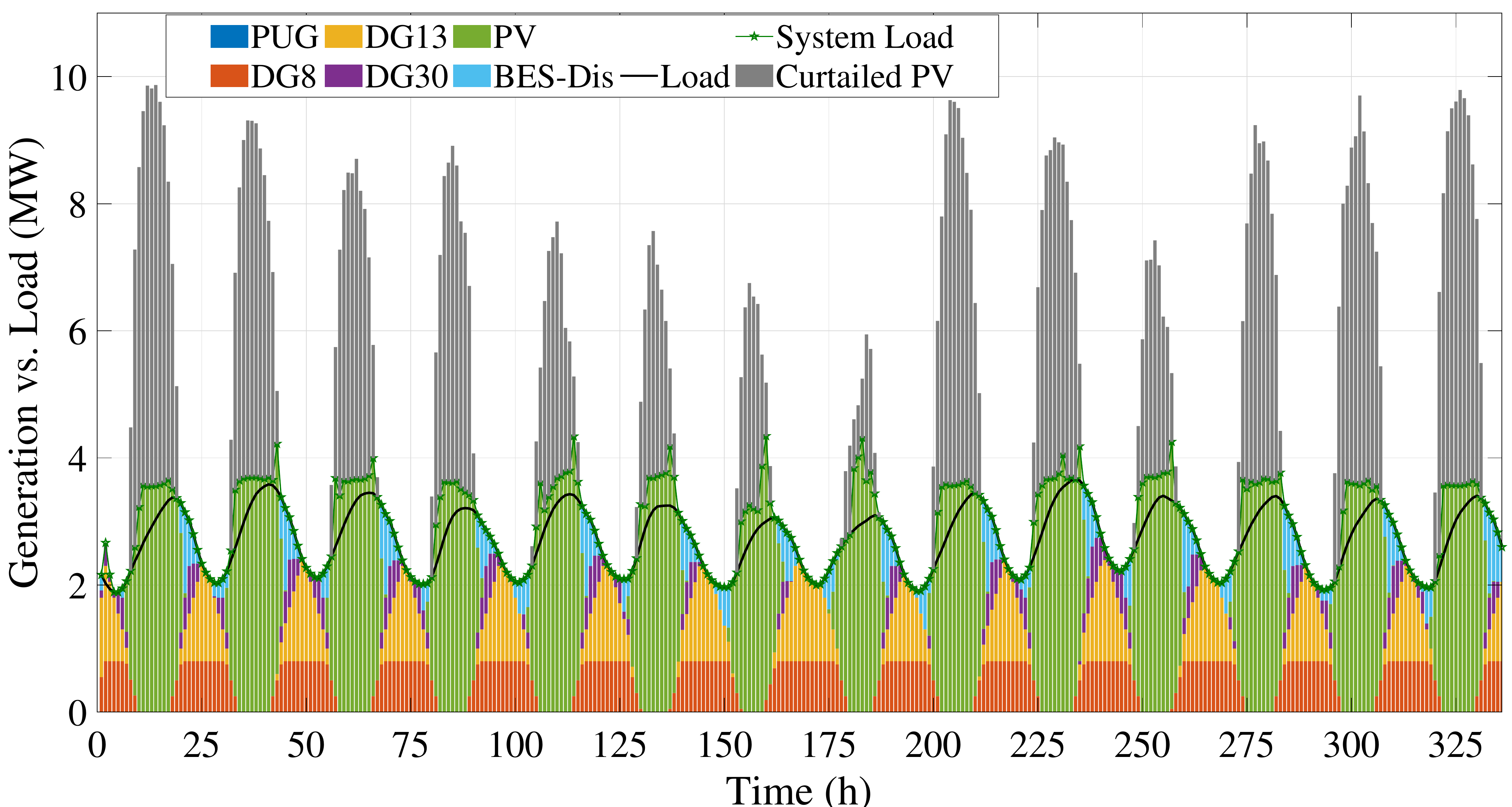}
	\caption{Optimal system operation based on DGs, PVs, and 4-hour duration battery (Case 3).}
	\label{Case3}
\end{figure}

In this case study, a 4-hour duration battery with round trip efficiency of 81\% is sized based on the year 2050 power and energy capital costs obtained from the National Renewable Energy Lab's ATB \cite{Vimmerstedt2018ATB}. Fig. \ref{Case3} shows that the battery starts charging when solar energy (green columns) that exceeds the grid load is available. The green columns above the black line (customer load) shows that the battery is charging and the total load of the system (shown as dark green line with stars) is increased over the black customer load line. The optimal power and energy sizes for the 4-hour duration battery are 1.44 MW and 5.76 MWh, respectively. At off-solar times the battery is discharging (shown with light blue columns) and injecting power to the grid lowering the use of DG30 energy.

\begin{figure}
\centering
\footnotesize
	\includegraphics[width=4.5in]{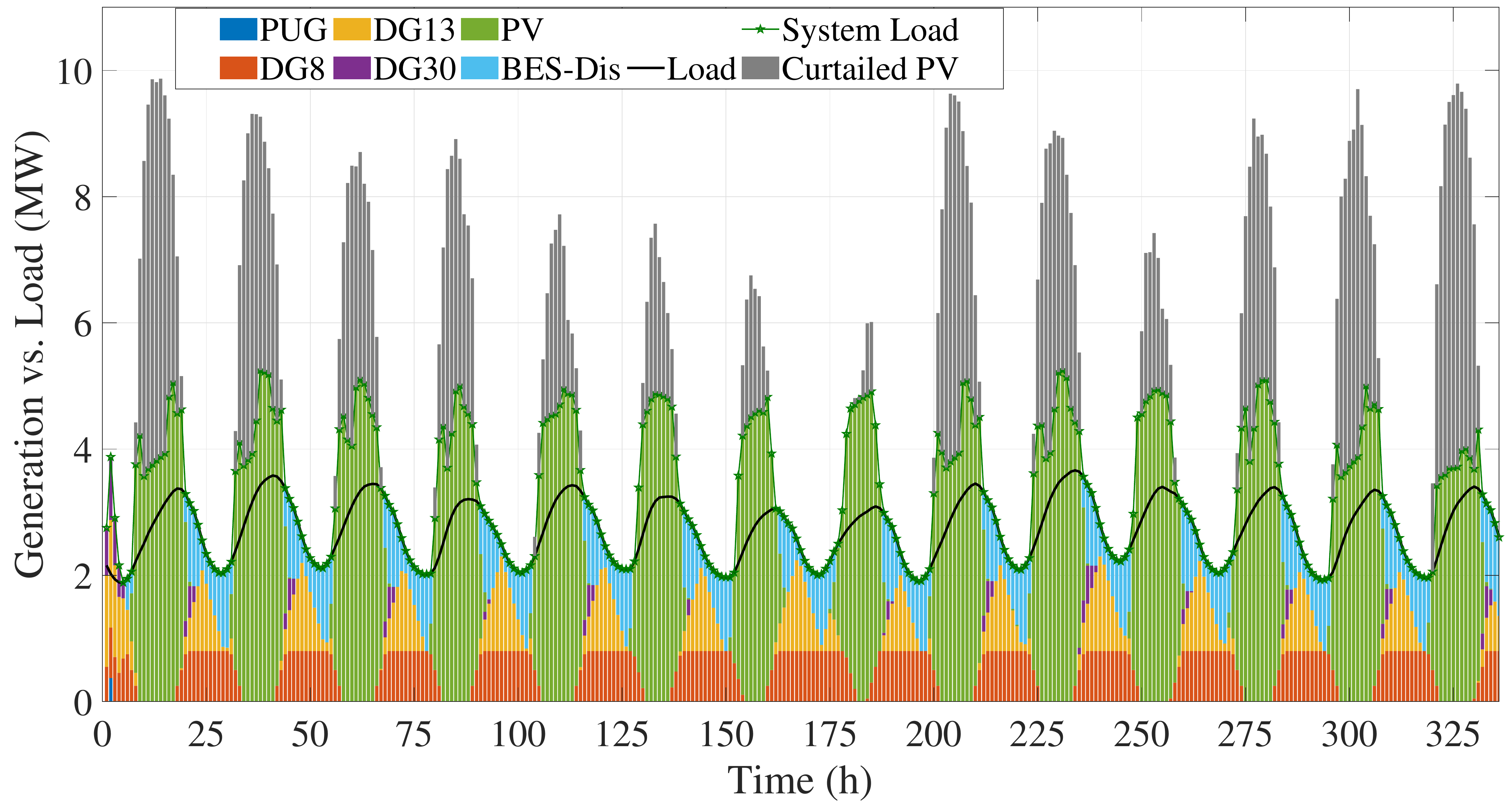}
	\caption{Optimal system operation based on DGs, PVs, and 10-hour duration redox flow battery (Case 4).}
	\label{Case4}
\end{figure}

\subsubsection{Optimal system operation based on DGs, PVs, and 10-hour duration redox flow battery (Case 4)} In this scenario, a 10-hour duration redox flow battery was added to the system of Case 2, at node 6 of Fig.\ref{system}. As with the 4-hour duration battery, the redox flow battery with the round trip efficiency of 67\% is sized based on the power and energy capital costs. The power and energy capacities are 1.52 MW and 15.2 MWh, respectively. Fig. \ref{Case4} shows that the 10-hour battery charges longer on solar energy as the green columns are further above the black line than in Fig. 4. Additionally, the 10-hour duration redox flow battery assists the grid at off-solar times almost eliminating the need of DG30. This case further increases the green energy production.

\subsubsection{Optimal system operation based on DGs, PVs, and long-term $H_2$ storage (Case 5a)} 
\begin{figure}
\centering
\footnotesize

	\includegraphics[width=4.5in]{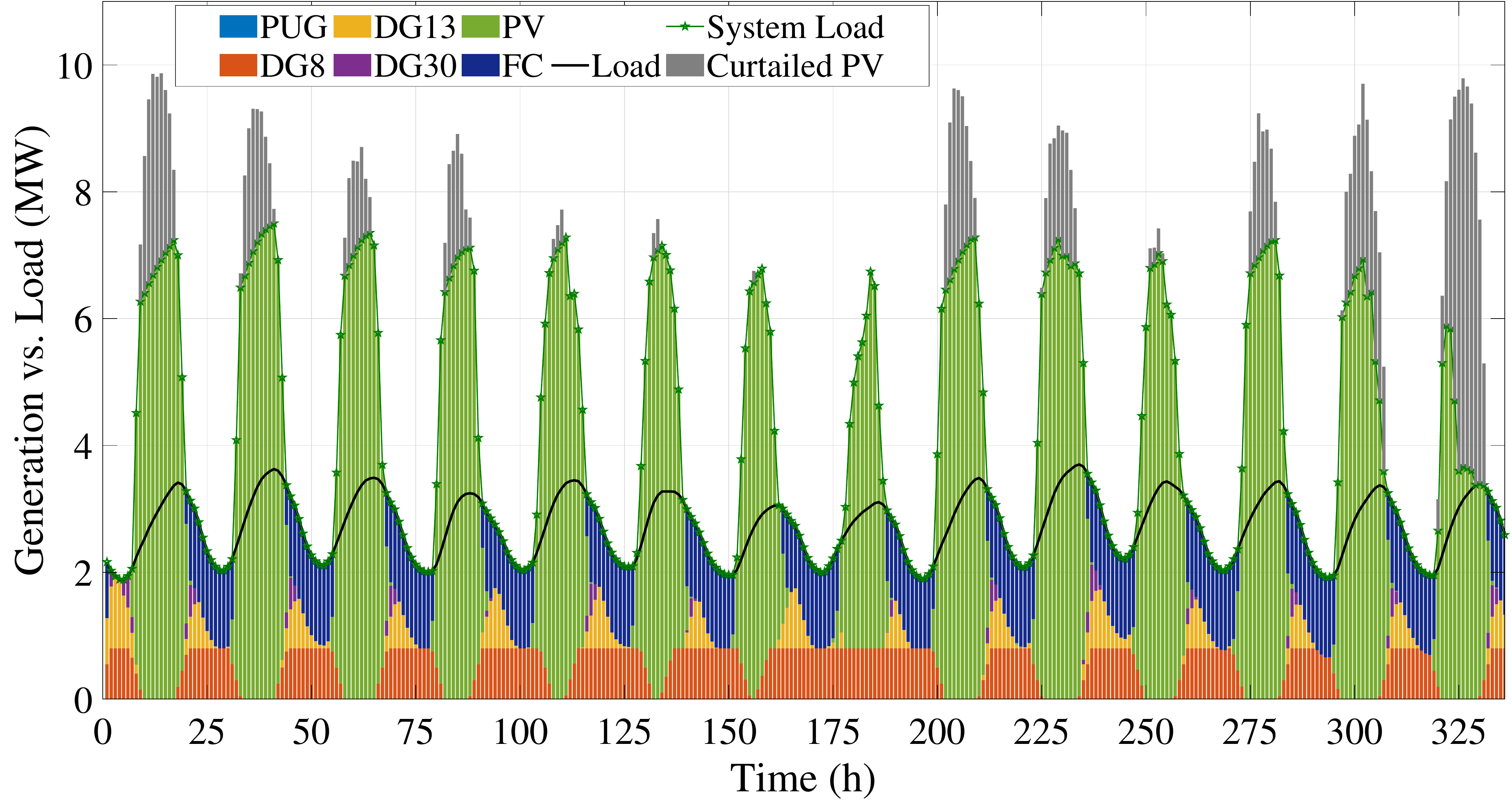}
	\caption{Optimal operation based on DGs, PVs, and $H_2$ system (Case 5a).}
	\label{Case5_updatedCF}
\end{figure}

Optimal operation of DGs and PV units with a $H_2$ system at node 19 was investigated in this case and is shown in Fig.\ref{Case5_updatedCF}. The optimum sizes for the electrolyzer (with 60\% efficiency), storage tank, and FC unit (with 70\% efficiency) are 3.87 MW, 1117.9 kg of $H_2$, and 1.26 MW respectively. The fuel cell (dark blue columns) was able to provide electricity for all of the evenings.  While the hydrogen storage is less efficient than the short term battery storage systems, very little of the PV electricity was curtailed.  For the fuel cell to inject 1 MWh of energy the electrolyzer must consume 2.38 MWh. As longer term storage of $H_2$ in a tank is available there is far less operation of DG30 (purple columns) and DG13 (orange columns) which results in much higher green energy consumption and carbon emissions are substantially reduced. In this case the initial amount of $H_2$ in the tank was 10\% and the final amount of $H_2$ in the tank was 10\% causing initial use of DG13 in the early hours of the two week period of study and curtailed PV at the end of the two week period.

\subsubsection{Optimal system operation based on DGs, PVs, and long-term $H_2$ storage with updating the LCOEs of DGs based on their actual capacity factor (Case 5b)} 

\begin{figure}
\centering
\footnotesize
	\includegraphics[width=4.5in]{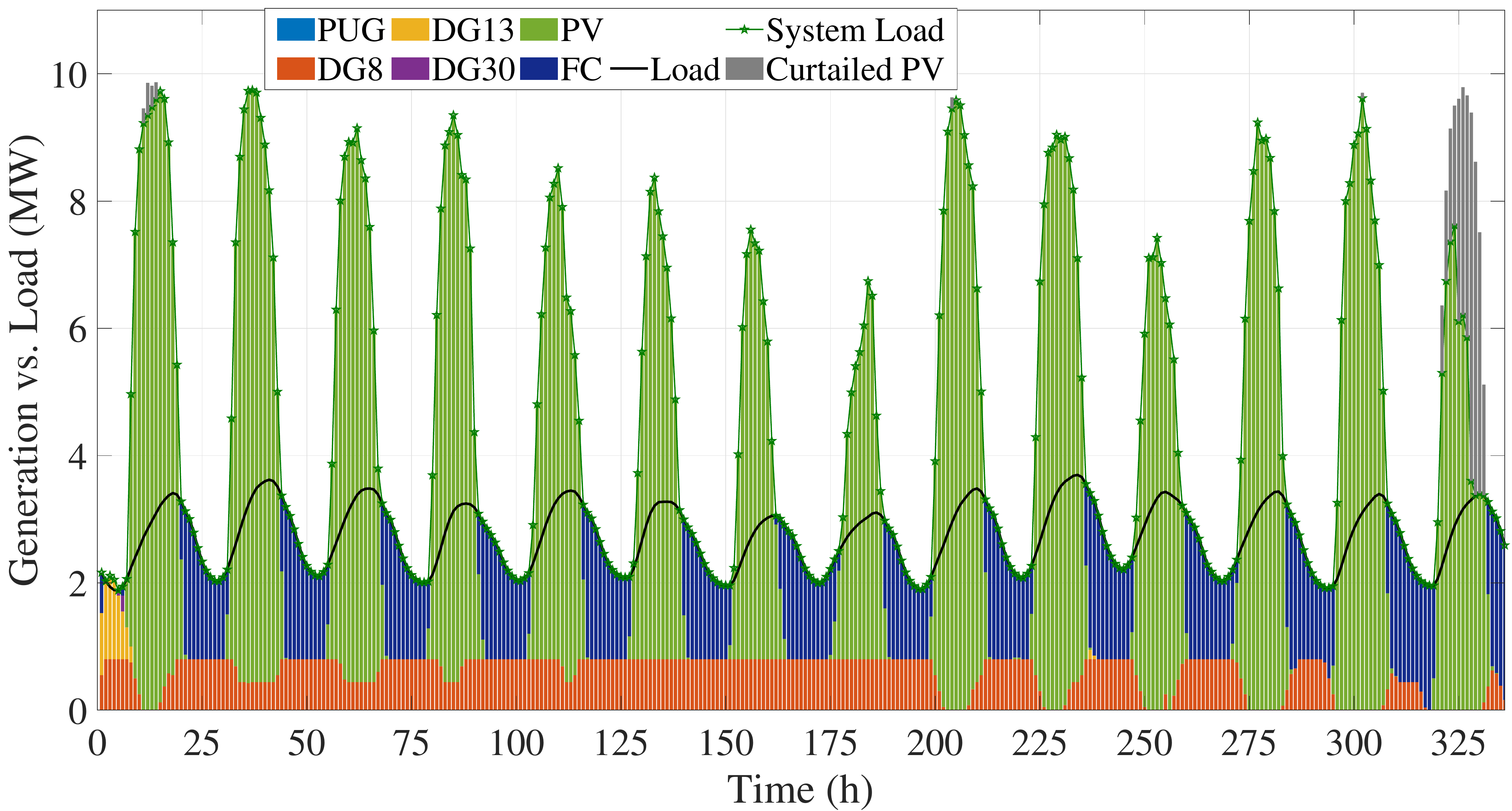}
	\caption{Optimal operation based on DGs, PVs, and $H_2$ system with updated LCOEs of DG units (Case 5b).}
	\label{Case5b}
\end{figure}

In Fig. \ref{Case5b}, the DGs' levelized cost of energy (LCOEs) at standard capacity factor are chosen based on the NREL ATB file \cite{Vimmerstedt2018ATB}. However, the optimum operation of these DERs significantly changes the capacity factors of all three DGs. For this reason, Fig. \ref{Case5b} shows the optimum operation of Case 5a considering actual DGs capacity factors and their corresponding LCOEs. For this case, optimum sizes of the electrolyzer, storage, and FC are 6.5 MW, 1934 kg of $H_2$, and 2.43 MW, respectively.

\begin{figure}
\centering
\footnotesize
	\includegraphics[width=4.5in]{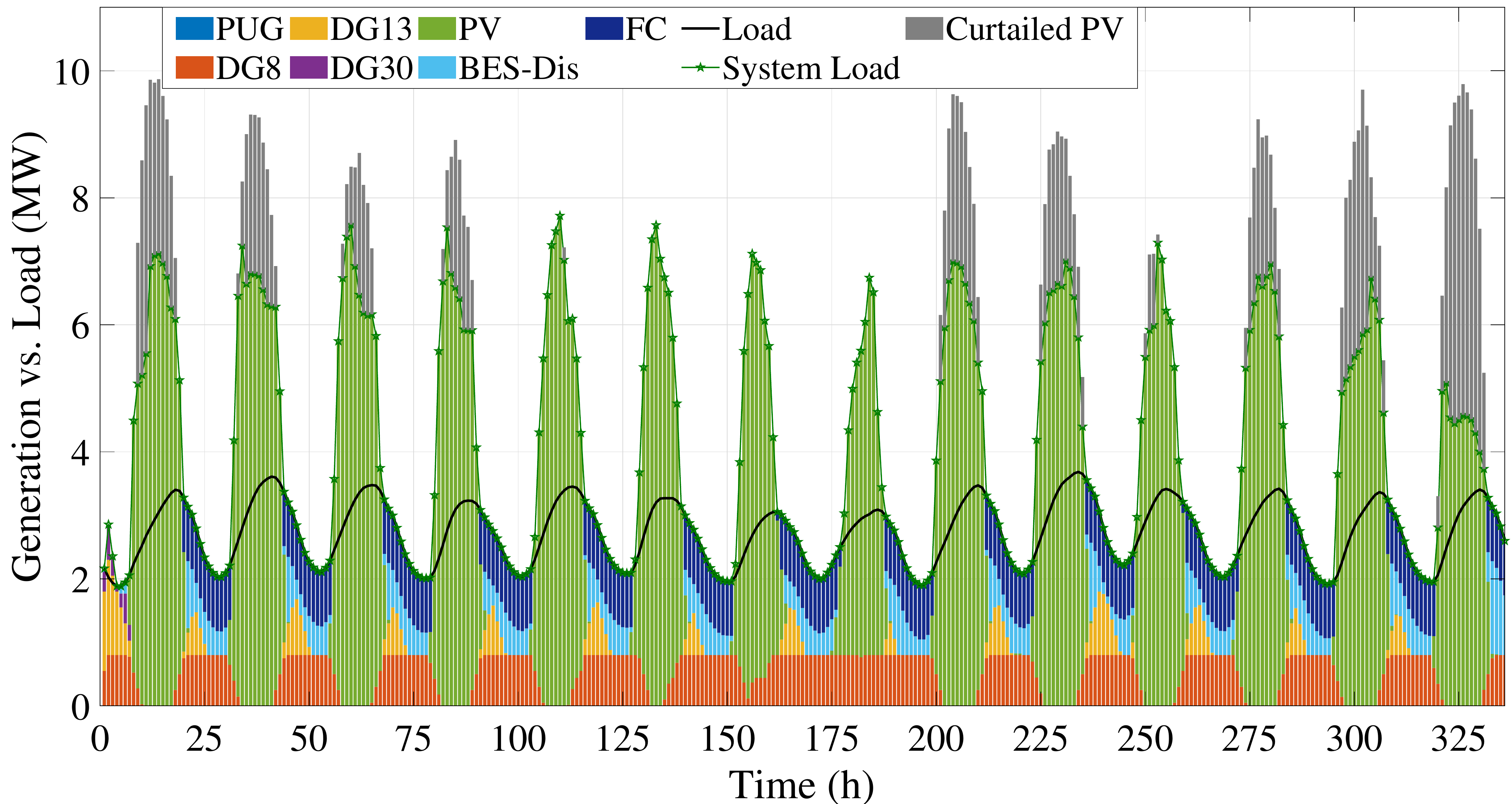}
	\caption{Results for optimal operation only based on DGs, PVs, $H_2$ system and Li-ion battery (Case 6a).}
	\label{Case6}
\end{figure}

\subsubsection{Optimal system operation based on DGs, PVs, 4-hour duration Li-ion battery and long-term $H_2$ storage (Case 6a)} In this case, a simultaneous sizing algorithm was implemented for the battery and $H_2$ system. Based on the results shown in Fig.\ref{Case6}, the battery power and energy ratings are 1.875 MW and 7.5 MWh and the $H_2$ system optimum sizes for electrolyzer, storage tank, and FC units are 2.69 MW, 823.6 kg of $H_2$, and 0.858 MW respectively. It should be noted that with both battery and hydrogen storage available, DG30 is almost eliminated from operation.

\subsubsection{Optimal operation of the 4-hour duration Li-ion battery and long-term $H_2$ storage while penalizing the curtailed PV energy in the objective function (Case 6b)} To maximize the green energy production and consumption, PV curtailment is penalized in the objective function. In this scenario, the otherwise curtailed PV energy is dispatched by the $H_2$ system and battery. Since the curtailed PV energy is penalized in the objective function, the optimal results for sizing of the assets are also changed. The electrolyzer, storage tank, and FC sizes are changed to 4.34 MW, 1323 kg of $H_2$, and 1.273 MW respectively. Additionally, battery power and energy ratings are changed to 1.354 MW and 5.42 MWh.  

\begin{figure}
\centering
\footnotesize

	\includegraphics[width=4.5in]{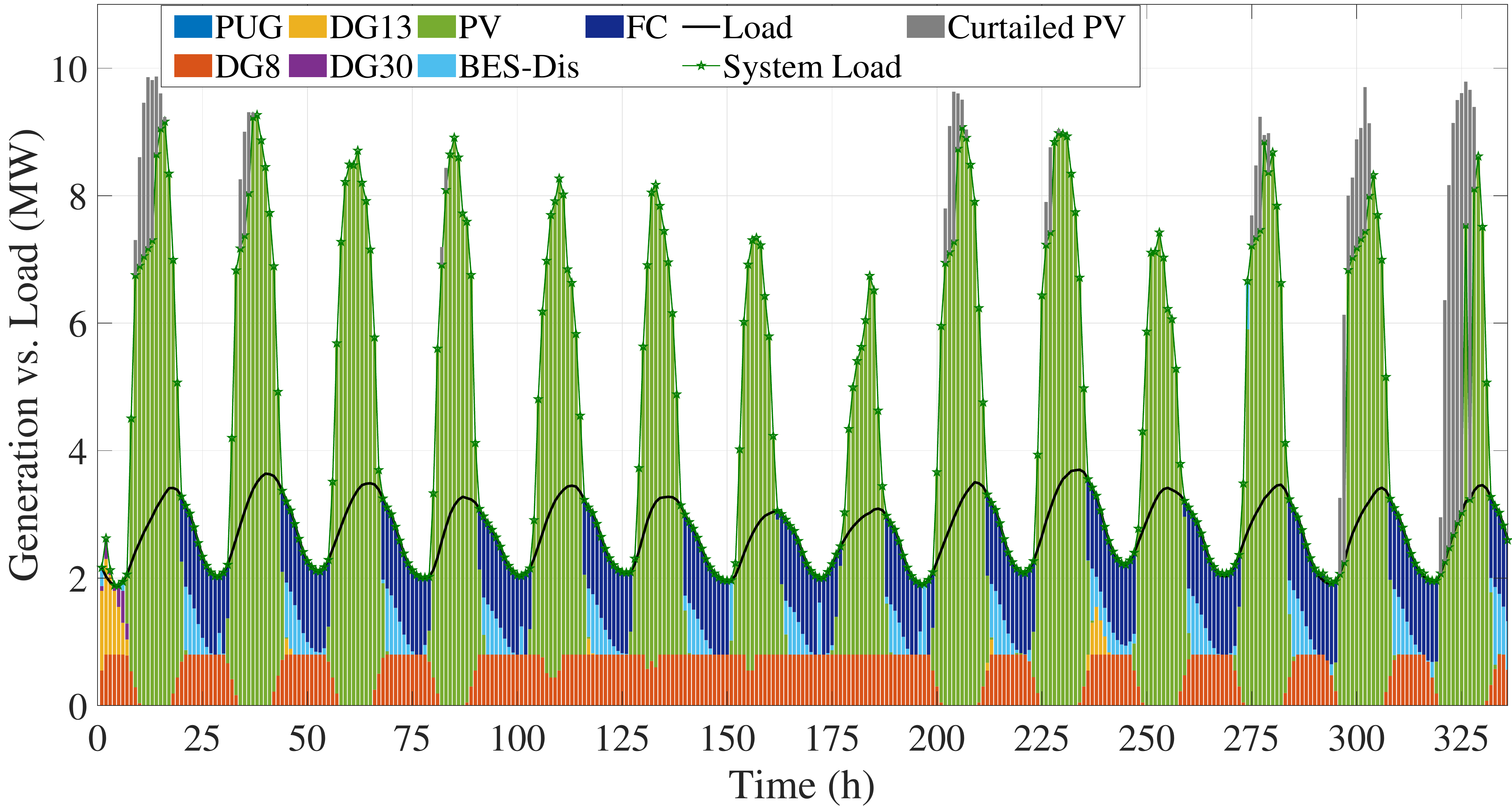}
	\caption{Results for optimal operation of case 6 considering penalizing the PV curtailment in the objective function (Case 6b).}
	\label{case6_penalized}
\end{figure}

\subsubsection{Optimal system operation based on DGs, PVs, 10-hour duration redox flow battery and long-term $H_2$ storage (Case 7a)}

\begin{figure}
\centering
\footnotesize

	\includegraphics[width=4.5in]{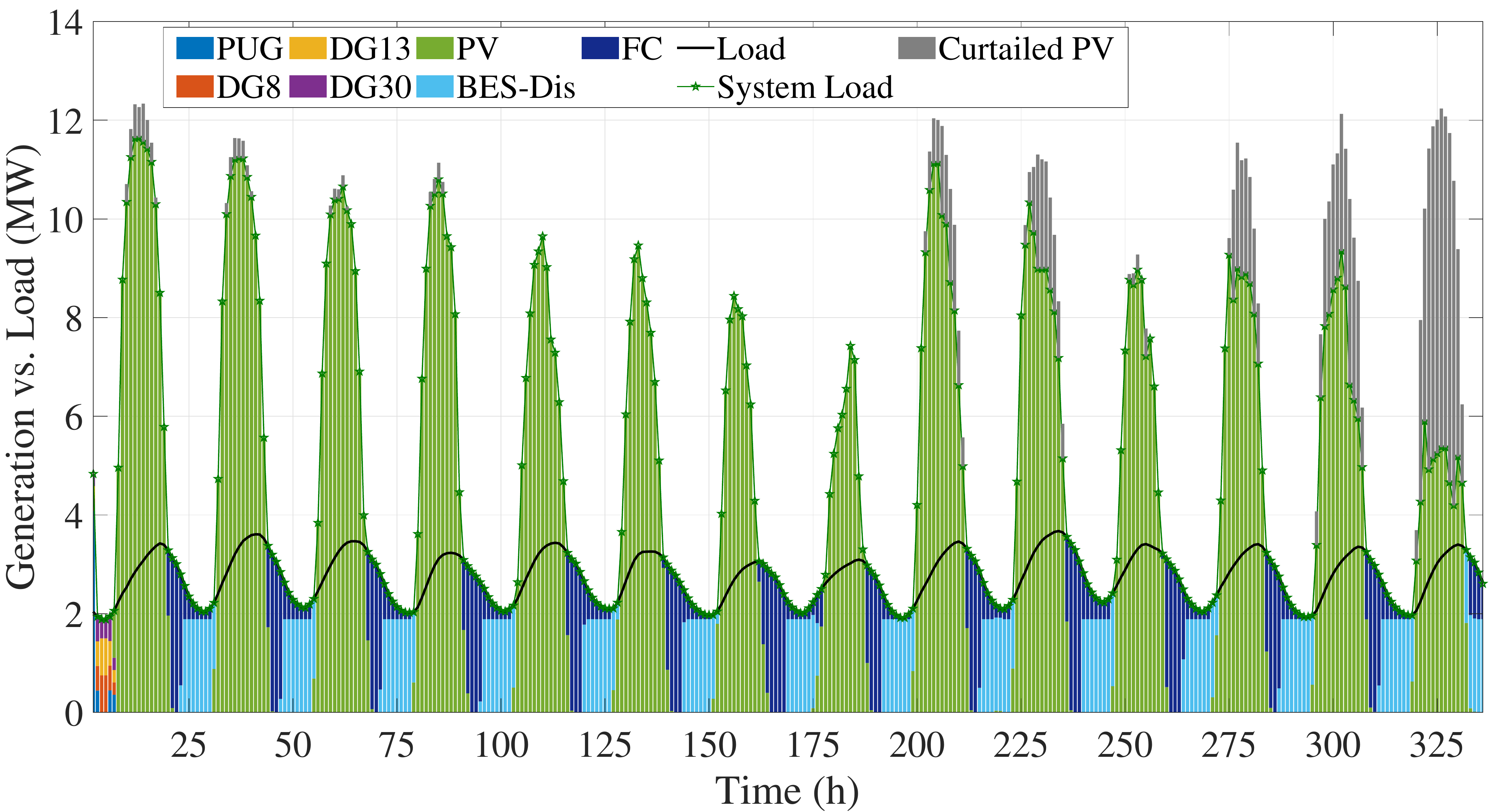}
	\caption{Results for optimal fossil-free operation of 10-hour duration flow battery and $H_2$ system for high PV penetration (150\% of grid load) (Case 7a).}
	\label{basecase8}
\end{figure}
Fig. \ref{basecase8} shows the optimal operation of a 10-hour duration redox flow battery and $H_2$ system. Due to the inefficiency of flow battery and $H_2$ system, the PV penetration was increased to 150\% of the total 33-node grid load (without considering the battery charging and electrolyzer load) to reach maximum green energy production. The total load is supplied by both the flow battery and $H_2$ system, and the green energy production in this case is 99.1\%. The 99.1\% green energy production is not 100\% as the three DGs must operate in the first hours of Day 1 as there is not enough $H_2$ available in the tank for the FC. By increasing the initial $H_2$ in the tank to 50\% of its capacity (which is shown in Fig. \ref{Fossilfree_initial50}), the green energy production is 100\% as the FC injected power during the first hours (this scenario is shown as Case 7b). The optimum sizes for battery power rating, electrolyzer, storage tank, and FC unit are 2.3 MW, 6.49 MW, 2109 kg of $H_2$, and 3.38 MW, respectively. 
  
\begin{figure}
\centering
\footnotesize

	\includegraphics[width=4.5in]{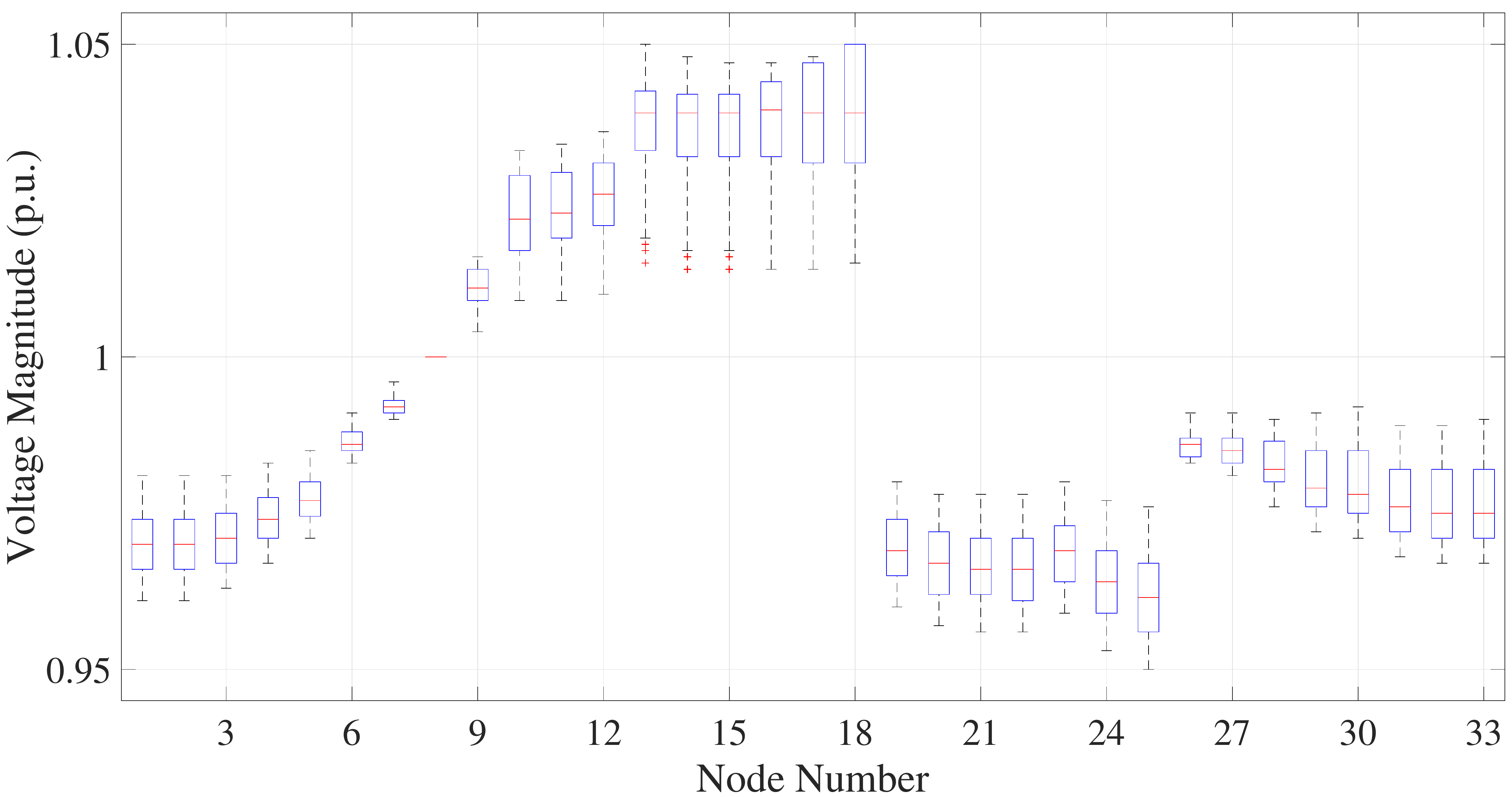}
	\caption{Voltage values for Case 2 (Operation based on DGs and PVs only).}
	\label{voltage_case2}
\end{figure}

\begin{figure}
\centering
\footnotesize

	\includegraphics[width=4.5in]{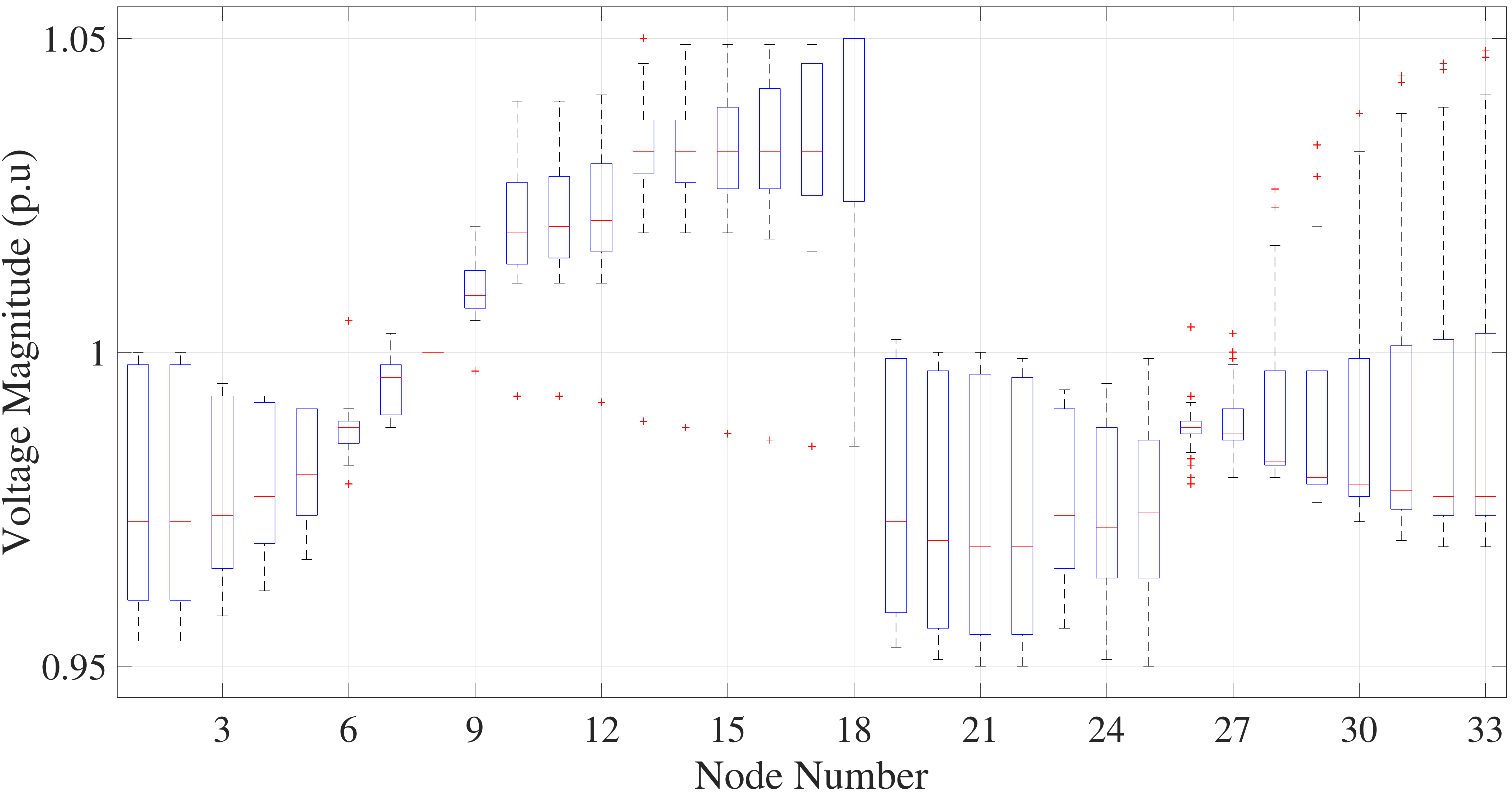}
	\caption{Voltage values for Case 7 (Fossil free scenario with $H_2$ storage and redox flow battery).}
	\label{basecase8_voltage}
\end{figure}

\begin{figure}
\centering
\footnotesize

	\includegraphics[width=4.5in]{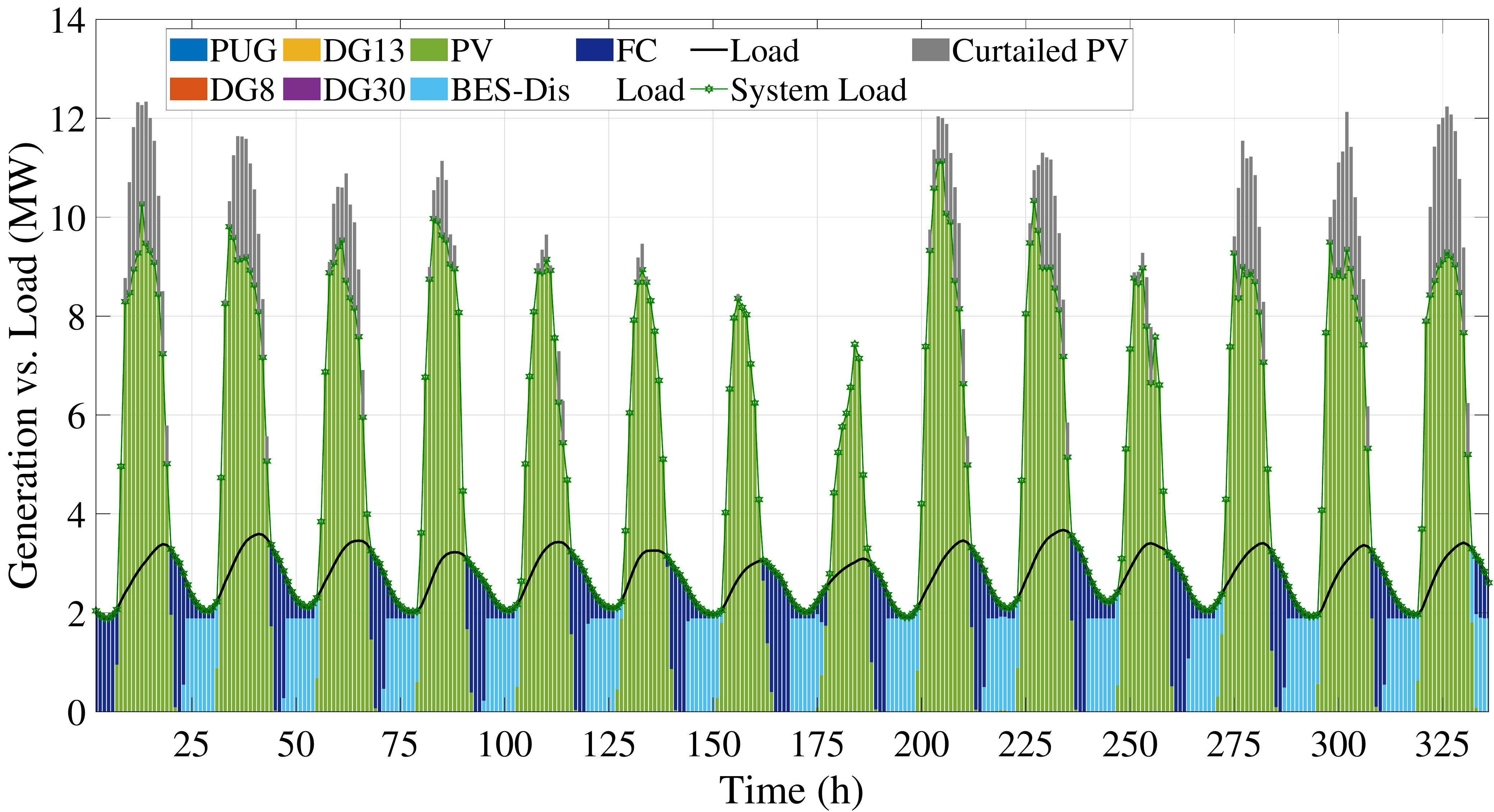}
	\caption{Results for optimal fossil-free operation of 10-hours duration flow battery and $H_2$ system for high PV penetration (150\% of grid load) considering 50\% initial and end level for $H_2$ mass (Case 7b).}
	\label{Fossilfree_initial50}
\end{figure}

Fig.  \ref{voltage_case2} and Fig.  \ref{basecase8_voltage} show that the voltage magnitudes at each node are within the specified limit defined by national American standards (i.e., $\pm 5\%$ of nominal voltage in p.u.)  \cite{kusko2007power} for the 100\% fossil fuel generation and the near 100\% green electricity generation case studies, respectively. Fig.  \ref{voltage_case2} is intentionally shown as a benchmark (which is related to the Case 2) for highlighting the role of DERs inverters in improving the voltage profile. The inverters on the solar, batteries and fuel cells improve the overall voltage profile of the 33-node system in comparison to fossil fuel generation. 
More details regarding the inverter/converter of DERs and their role in power flow control in normal and emergency operations can be found in \cite{hayerikhiyavi2021comprehensive}.

\subsubsection{Analysis of DGs' capacity factor for different case studies}
\begin{figure}
\centering
\footnotesize

	\includegraphics[width=4.5in]{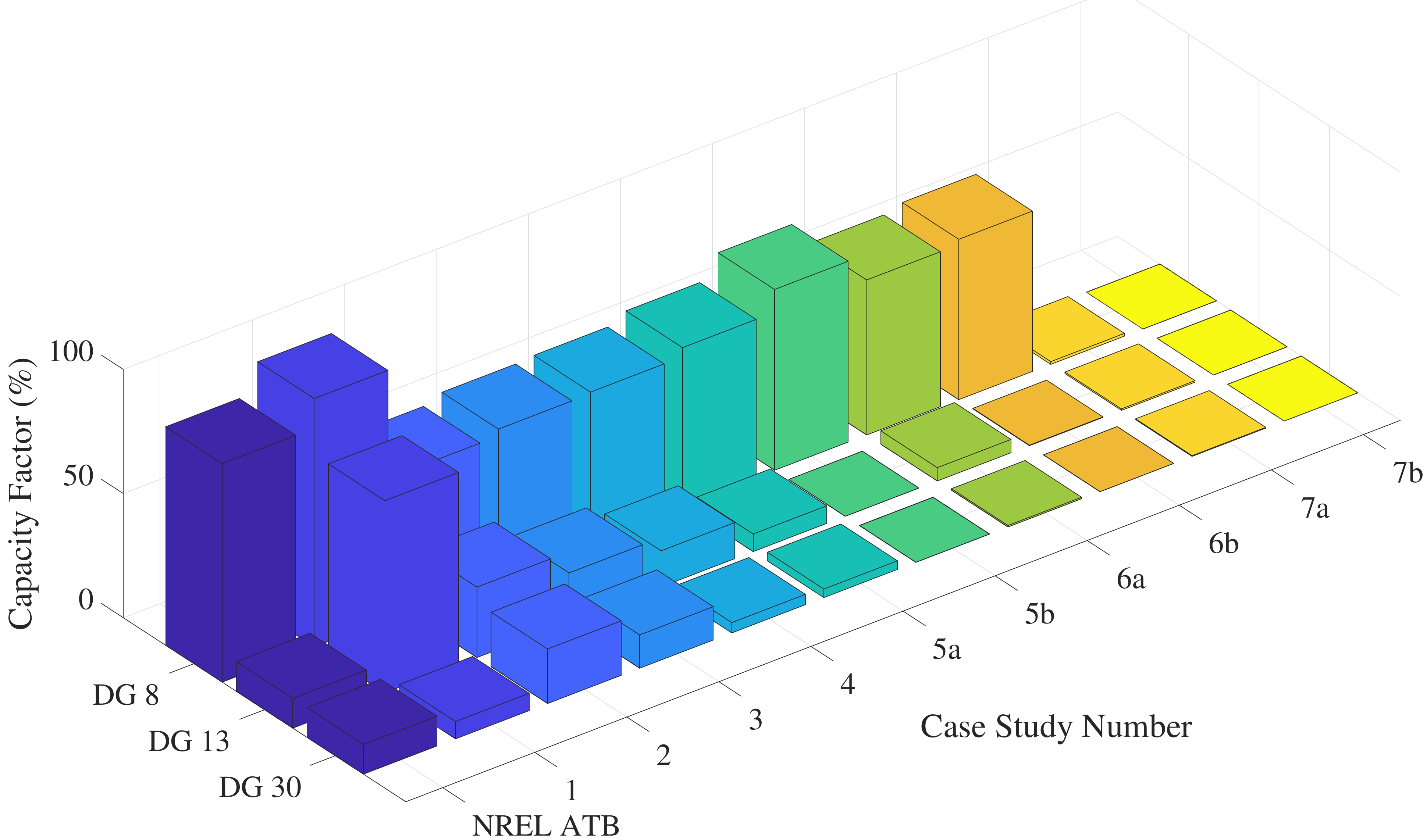}
	\caption{Comparison of DG's capacity factor for different case studies.}
	\label{DGcf}
\end{figure}

Fig. \ref{DGcf} shows the capacity factor of the DGs for the different case studies. The first three columns (for DG 8, DG 13, and DG 30) show the NREL ATB's \cite{Vimmerstedt2018ATB} reference capacity factors that were used as benchmarks. In Case 1, both DG 8 and DG 13 actual capacity factors are higher than the benchmark and the DG 30 capacity factor is lower than NREL ATB benchmark. In Case 2, PV units were added to the system and some portion of the grid load is supplied by PV energy, reducing the capacity factor of DG 8 and DG 13. The  addition of a large amount of PV energy leads to ramp rate issues for all three of the DGs, causing DG 30, a peaking unit, to operate for more hours compared to the previous cases. From Case 3 to Case 7 b, as more distributed PV and energy storage were added the capacity factors of the DGs were reduced. In Case 5b, the assets were sized based on the actual capacity factor and then their actual LCOEs, which were calculated based on the following equation:

\begin{equation}
    Actual \; LCOE = \frac{NREL \;Capacity \;Factor}{Actual \;Capacity \;Factor} \times NREL \;LCOE
\end{equation}

Compared to the benchmark capacity factors and LCOEs, updating LCOEs in the objective function based on the actual lower capacity factors, results in very high LCOEs. This is clearly shown in Case 5a and Case 5b in which the LCOEs of Case 5a are updated based on DGs actual capacity factors and results were presented in Case 5b. Considering Fig. \ref{DGcf}, as the LCOEs were updated (Case 5b), DG 13 and DG30 capacity factors were reduced compared to Case 5a, and DG 8 capacity factor was increased to be close enough to the DG 8 capacity factor presented in NREL ATB's file. As the DGs capacity factors changed, the size of the electrolyzer, storage tank, and FC increased to replace the most expensive DGs (DG 13 and DG 30 with their actual LCOEs based on their actual capacity factor). In Case 7a and Case 7b, DG 13 and DG 30 capacity factors are less than 0.1\% and as a consequence their actual LCOE are \$11,400 /MWh. As the 10-hour duration battery and $H_2$ system cost-effectively supply the load. More green energy is produced and there is close to net-zero emission electricity production. 

\subsubsection{Analysis of green energy production and DLMP for different case studies}
\begin{figure}
\centering
\footnotesize

	\includegraphics[width=4.5in]{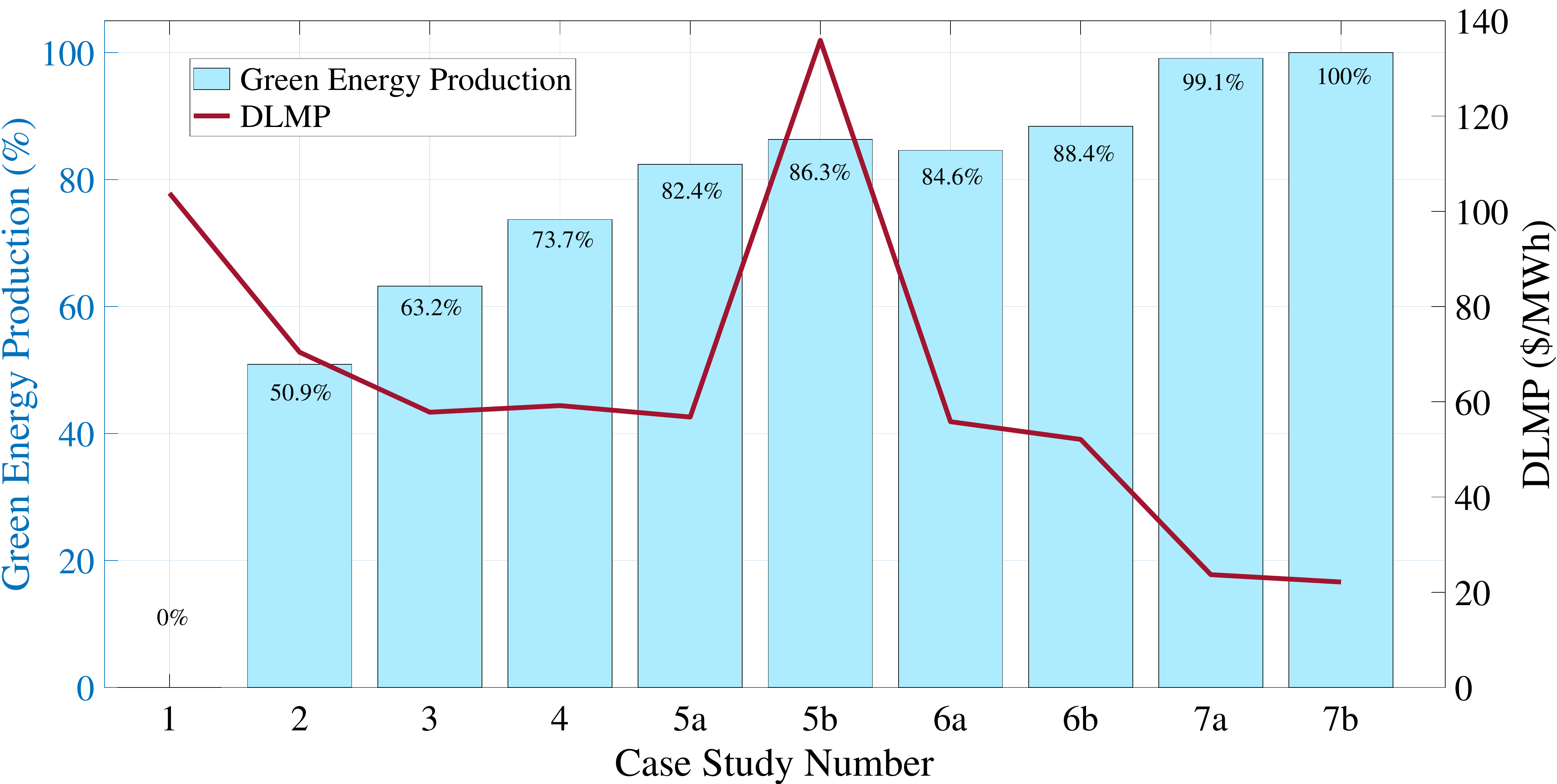}
	\caption{Green energy production and DLMP for different case studies.}
	\label{greenEnergy}
\end{figure}

The Green energy production (left-side y-axis) and DLMP (right-side y-axis) curves are shown in Fig. \ref{greenEnergy} for all different case studies. The green energy production for Case 1 is zero, since only fossil fuel DGs are operating. As PV units, battery storage technologies, and $H_2$ systems are added, the green energy production increases gradually to 100\% (The maximum green energy production occurs in Case 7b where there is enough $H_2$ in the tank initially and there is ten hours of battery storage. As the green energy increases, the DLMP price decreases from \$103/MWh to \$22/MWh, except for Case 5b. In Case 5b, since the LCOE of the DGs are updated based on their actual capacity factor compared to the NREL ATBs' reference values, the higher LCOEs of the DGs result in both higher DLMP values and more green energy production. The minimum DLMP occurs in Case 7b where all three DGs are off and the PV, $H_2$  system, and redox flow battery are responsible for supplying the customer electricity demand.

\begin{table}[]
\centering
\caption{Total PV energy capacity and curtailed PV energy during two weeks period for different case studies}
\label{table}
\begin{tabular}{|c|c|ccccccc|cc|}
\hline
Case &
  1 &
  \multicolumn{1}{c|}{2} &
  \multicolumn{1}{c|}{3} &
  \multicolumn{1}{c|}{4} &
  \multicolumn{1}{c|}{5a} &
  \multicolumn{1}{c|}{5b} &
  \multicolumn{1}{c|}{6a} &
  6b &
  \multicolumn{1}{c|}{7a} &
  7b \\ \hline
\begin{tabular}[c]{@{}c@{}}Total PV\\  Energy \\ (MWh)\end{tabular} &
  0 &
  \multicolumn{7}{c|}{1110.8} &
  \multicolumn{2}{c|}{1388.5} \\ \hline
\begin{tabular}[c]{@{}c@{}}Total \\ Curtailed \\ PV\\ Energy\\  (MWh)\end{tabular} &
  0 &
  \multicolumn{1}{c|}{649.1} &
  \multicolumn{1}{c|}{545.5} &
  \multicolumn{1}{c|}{418.7} &
  \multicolumn{1}{c|}{161.3} &
  \multicolumn{1}{c|}{35.5} &
  \multicolumn{1}{c|}{194} &
  83.6 &
  \multicolumn{1}{c|}{139.1} &
  146.2 \\ \hline
\begin{tabular}[c]{@{}c@{}}Total \\ Curtailed \\ PV\\ Energy\\  (\%)\end{tabular} &
  0 &
  \multicolumn{1}{c|}{58.4} &
  \multicolumn{1}{c|}{49.1} &
  \multicolumn{1}{c|}{37.7} &
  \multicolumn{1}{c|}{14.5} &
  \multicolumn{1}{c|}{3.2} &
  \multicolumn{1}{c|}{17.5} &
  7.5 &
  \multicolumn{1}{c|}{10} &
  10.5\\ \hline
\end{tabular}
\end{table}

\subsubsection{Extra analysis of case studies}
Table \ref{table} shows the total energy of the six PV units produced over two week period, and the curtailed solar energy for the different case studies. In Cases 1 to 6, the total PV energy is 120\% of the customer grid load, 1110.8 MWh. In these cases, the minimum and maximum PV curtailment occurs in Case 5b (3.2\%) and Case 2 (58.4\%), respectively. Since there is no energy storage device in Case 2, PV energy only supplies the electricity demand of customers during the time when there is abundant solar energy and the rest is curtailed. In Case 5b, since the LCOE of DGs are more expensive than Case 5a the larger electrolyzer, storage tank, and FC use more of the PV energy leading to little solar energy curtailment. The electrolyzers have enough capacity to consume more PV energy, the storage tank is larger to store the additional $H_2$, and the fuel cell provides the energy to replace the DG energy.  Since the round trip energy efficiency of the $H_2$ system is 42\%, more PV energy must consumed to provide electricity at non-solar times when compared to the case studies with only the Li-ion battery (with 81\% round trip efficiency) or the redox flow battery (with 67\% round trip efficiency) as a result for the systems with only one storage system Case 5b. The lower round trip efficiencies of the flow battery and hydrogen storage systems, required that the PV energy be increased to 150\% of the customer grid load (1388.5 MWh) in Case 7a and Case 7b. Case 7b curtailed PV energy (10.5\%) is more than Case 7a (10\%), as the initial and final $H_2$ in the tank was increased from 10\% to 50\%.

\par


\begin{table}[]
\centering
\caption{Case 7 $H_2$ Production Cost Sensitivity Analysis Subject to Electrolyzer CAPEX Cost (PV LCOE at \$12/MWh)}
\label{CAPEX_Sensitivity}
\begin{tabular}{|c|c|c|c|c|c|c|c|c|}
\hline
\begin{tabular}[c]{@{}c@{}}Electrolyzer Capex Cost\\ (\$/kW)\end{tabular} & 50   & 75   & 100  & 125 & 150  & 175  & 200  & 250  \\ \hline
\begin{tabular}[c]{@{}c@{}}$H_2$ Production Cost \\ (\$/kg)\end{tabular}     & 0.85 & 0.94 & 1.02 & 1.1 & 1.19 & 1.27 & 1.35 & 1.51 \\ \hline
\begin{tabular}[c]{@{}c@{}}$H_2$ Storage Cost \\ (\$/kg)\end{tabular}     & 0.02 & 0.02 & 0.02 & 0.02 & 0.02& 0.02 & 0.02 & 0.02 \\ \hline
\begin{tabular}[c]{@{}c@{}} Total $H_2$ Cost \\ (\$/kg)\end{tabular}     & 0.87 & 0.96 & 1.04 & 1.12 & 1.21 & 1.29 & 1.37 & 1.53 \\ \hline
\end{tabular}
\end{table}

\begin{table}[]
\centering
\caption{Case 7 $H_2$ Production Cost Sensitivity Analysis Subject to Different PV LCOEs}
\label{H2cost_PVsensitivity}
\begin{tabular}{|c|c|c|c|c|c|c|}
\hline
\begin{tabular}[c]{@{}c@{}}PV LCOE (\$/MWh)\end{tabular}                                                     & 8    & 9    & 10  & 11   & 12   & 13   \\ \hline
\begin{tabular}[c]{@{}c@{}}$H_2$ Production Cost (\$/kg) for Electrolyzer \\ CAPEX Cost of 100 (\$/kW) with \\ Low Pressure (40 bar)\end{tabular} & 0.79 & 0.85 & 0.9 & 0.96 & 1.02 & 1.08 \\ \hline
\begin{tabular}[c]{@{}c@{}}$H_2$ Production Cost (\$/kg) for Electrolyzer\\ and Compressor CAPEX Cost of 248 (\$/kW) with \\ Moderate Pressure (350)\end{tabular}  & 1.28 & 1.34 & 1.4 & 1.46 & 1.51 & 1.56 \\ \hline
\end{tabular}
\end{table}

Generally, $H_2$ production cost depends on the CAPEX cost, capacity factor, and hourly energy price that the electrolyzer consumes. The $H_2$ electrolyzer production and storage cost are shown in Table 2 as a function of the electrolyzer CAPEX as applied to Case 7 (the values used for the earlier analysis are in the column highlighted with grey color). In Case 7, studied earlier the electrolyzer and compressor CAPEX costs were \$100/kW and \$148/kW, respectively. The PV LCOE in all cases of Table 2 was at \$12 /MWh. The cost of $H_2$ out of the electrolyzer with a CAPEX of \$100/kW is \$1.02/kg at the electrolyzer output of 40 bar. However, if the compression cost is also included, the $H_2$ cost out of the compressor will be approximately \$1.51 /kg. Table 2 also shows that the storage cost is \$0.02 /kg as constant number since it is not related to the electrolyzer CAPEX cost. To have a $H_2$ production cost lower than \$ 1 /kg (out of the electrolyzer), the electrolyzer CAPEX cost must be below \$ 100/kW at a PV LCOE of \$12/MWh.

Table \ref{H2cost_PVsensitivity} demonstrates the results for $H_2$ production cost subject to different PV LCOEs for stationary fuel cell operation, as done for this study at 40 bar, and for 350 bar as would be used to fuel medium to heavy duty fuel cell vehicles. The highlighted column shows the PV LCOE that originally was used for Case 7 simulations. If the CAPEX cost and consequently PV LCOE decreases significantly, the $H_2$ cost directly from the electrolyzer would be less than \$1/kg. Although the scope of this paper was to investigate the role of $H_2$ storage for improved utility operations (not supplying the transportation sector demand) by the year 2050, the $H_2$ cost by considering both electrolyzer and compressor CAPEX costs (\$100 /kW + \$148/kW = \$248 /kW) for moderate pressure (350 bar) application such as supplying the $H_2$ demand for fuel cell electric vehicles was also considered. Table \ref{H2cost_PVsensitivity} shows that for PV LCOE of \$12/MWh, produced and compressed $H_2$ can be produced at \$1.51/kg. Recently, the U.S. Department of Energy (DOE) announced that the CAPEX cost of electrolyzer must be reduced by 80\% from 2020's value(to \$100/kW), and the electricity price must be below \$20/MWh by the year 2030 to produce $H_2$ at DOE's desired cost of \$1/kg by 2030. All the LCOEs and CAPEX costs which were previously presented for the year 2050 can be used for the year 2030 if the DOE targets are met.

\section{Conclusion}
In this paper, a path to net zero emission energy production using $H_2$ and different battery technologies such as Li-ion and redox flow was presented from the DSO perspective who manages these assets to minimize the total investment and operation cost of the network while increasing green energy production. Several case studies were considered to show how the DSO can move down the path to 100\% green energy production while addressing the technical and physical constraints of the 33-node distribution network. The $H_2$ system and redox flow battery together ca provide 100\% green energy production. The sensitivity of utility PV LCOE and electrolyzer CAPEX costs on producing hydrogen at \$1/kg were presented. The DSO could provide 100\% renewable electricity and $H_2$ could be produced at a cost of \$1/kg by 2050 with conservative cost estimates and by 2030 with accelerated cost declines based on the U.S. DOE targets.

\section{Acknowledgement}
This work is supported by U.S. Department of Energy's award under grant DE-EE0008851.

\section{Appendix}
The following Table \ref{appendix} show the input cost parameters used for the optimization.

\begin{table}[h]
\centering
\caption{Input Cost Related parameters for Optimizing Energy Production}
\label{appendix}
\begin{tabular}{|l|c|c|c|c|l|l|}
\hline
\multicolumn{1}{|c|}{Asset} &
  \begin{tabular}[c]{@{}c@{}}DG8\\ LCOE \\ (\$/MWh)\end{tabular} &
  \begin{tabular}[c]{@{}c@{}}DG13\\ LCOE \\  (\$/MWh)\end{tabular} &
  \begin{tabular}[c]{@{}c@{}}DG30\\ LCOE \\ (\$/MWh)\end{tabular} &
  \begin{tabular}[c]{@{}c@{}}PV\\ LCOE \\ (\$/MWh)\end{tabular} &
  \multicolumn{1}{c|}{\begin{tabular}[c]{@{}c@{}}Li-ion Battery\\ CAPEX\\ (\$/kW)\end{tabular}} &
  \multicolumn{1}{c|}{\begin{tabular}[c]{@{}c@{}}Flow Battery\\ CAPEX\\ (\$/kW)\end{tabular}} \\ \hline \hline
\multicolumn{1}{|c|}{Value} & 36  & 95  & 98  & 12  & \multicolumn{1}{c|}{613} & \multicolumn{1}{c|}{1370} \\ \hline \hline
Asset &
  \begin{tabular}[c]{@{}c@{}}Electrolyzer\\ CAPEX\\ (\$/kW)\end{tabular} &
  \begin{tabular}[c]{@{}c@{}}Compressor \\ CAPEX\\ (\$/kW)\end{tabular} &
  \begin{tabular}[c]{@{}c@{}}Storage \\ Tank \\ CAPEX\\ (\$/kg)\end{tabular} &
  \begin{tabular}[c]{@{}c@{}}Fuel Cell\\ CAPEX\\ (\$/kW)\end{tabular} &
   &
   \\ \hline \hline
Value                       & 100 & 148 & 240 & 500 &                          &                           \\ \hline 
\end{tabular}
\end{table}


%

\ifCLASSOPTIONcaptionsoff
  \newpage
\fi

\bibliographystyle{IEEEtran}
\bibliography{main}

\end{document}